\documentclass[showpacs,aps,amssymb,floatfix,prd,preprintnumbers]{revtex4-2}
\usepackage{epstopdf}
\usepackage{capt-of}
\usepackage{graphicx}  
\usepackage{dcolumn}   
\usepackage{bm}
\usepackage{amsmath}
\usepackage[font=scriptsize]{caption}
\usepackage[colorlinks]{hyperref}
\begin{document}
\title{\bf Modeling of accelerating Universe with bulk viscous fluid in Bianchi V space-time}

\author{G. K. Goswami \footnote{Department of Mathematics, Netaji Subhas University of Technology, Delhi, India, E-mail: gk.goswami09@gmail.com}, Anil Kumar Yadav \footnote{Department of Physics, United College of Engineering and Research, Greater Noida – 201 306, India, E-mail: abanilyadav@yahoo.co.in}, B. Mishra \footnote{Department of Mathematics, Birla Institute of Technology and Science-Pilani, Hyderabad Campus, Hyderabad-500078, India, E-mail: bivudutta@yahoo.com}, S. K. Tripathy \footnote{Department of Physics, Indira Gandhi Institute of Technology, Sarang, Dhenkanal-759146, Odisha, India, E-mail: tripathy\_sunil@rediffmail.com}}

\begin{abstract}
In this paper, we have investigated a bulk viscous anisotropic Universe and constrained its model parameters with recent $H(z)$ and Pantheon compilation data. Using cosmic chronometric technique, we estimate the present value of Hubble's constant as $H_{0} = 69.39 \pm 1.54~km~s^{-1}Mpc^{-1}$, $70.016 \pm 1.65~km~s^{-1}Mpc^{-1}$ and $69.36 \pm 1.42~km~s^{-1}Mpc^{-1}$ by bounding our derived model with recent $H(z)$ data, Pantheon and joint $H(z)$ and Pantheon data respectively. The present age of the Universe is specified as $t_0= 0.9796H_0^{-1}\sim 13.79$ Gyrs. The model favours a transitioning Universe with the transition red-shift as $z_{t} = 0.73$. We have reconstructed the jerk parameter using the observational data sets. From the analysis of the jerk parameter, it is observed that, our derived model shows a marginal departure from the concordance $\Lambda$CDM model. 

\end{abstract}



\maketitle

\section{Introduction} 
Recent cosmological observations have already approved the accelerated expansion of the Universe at least in the present epoch \cite{Riess98, Perlmutter99}. It has been cited that an unknown form of energy with negative pressure may be the reason behind the accelerated expansion of the Universe. Though the exact form of this energy is still unknown, the recent observational results indicate that it  occupies 68.3 percent of the mass energy budget of the Universe \cite{Ade16}. This exotic form of energy is  termed as dark energy which is believed to be less effective in early times but dominates at the present epoch. One intriguing fact of this dark energy is that, it does not interact with the baryonic matter and therefore it is difficult to  detect such an exotic dark energy form. Theoretically, this can be studied with the cosmic expansion history $H(z)$ and the growth rate of cosmic large scale structure $f_g(z)$ \cite{Ma99}. Several theoretical models based on the cosmic expansion phenomena have been proposed in last two decades; mostly post supernovae observations. Sahni and Shtanov\cite{Sahni03} have presented a new class of braneworld models, in which the scalar curvature of the brane metric contributes to the brane action. Based on the Karolyhazy relation, Wei and Cai \cite{Wei08} proposed the agegraphic dark energy model for a Friedmann-Robertson-Walker (FRW) Universe. Caplar and Stefancic \cite{Caplar13} have framed a generalized models of unification of dark matter and dark energy, and with the observational data on Hubble parameter at different redshifts, constrained the model parameters. Mishra and Tripathy \cite{Mishra15} have suggested a hybrid scale factor which can mimic the cosmic transition from early decelerated phase to accelerated phase at late time. Several such scale factors have been investigated to mimic this transitive behavior \cite{Mishra17, Mishra18a}. Farnes \cite{Farnes18} has suggested a toy model that leads to a cyclic Universe with a time variable Hubble parameter. The model provides compatibility with the current tension, which has been emerged in the cosmological measurements. Mishra et al. \cite{Mishra18b} have found the anisotropy in the dark energy pressure have evolved at late time. The cosmological models in a non-interacting two fluid scenario such as the usual dark energy and electromagnetic field  have been studied and have shown the dominance of dark energy at the late time of the evolution \cite{Ray19,Mishra19a}.  Some useful applications of dark components in anisotropic space time are given in Refs. \cite{Yadav/2011, Kumar/2011mpla, Amirhashchi/2011plb, Yadav/2012, Goswami/2015, Goswami/2016, Goswami/2016a, Goswami/2016b, Singla/2020, SKT2015, SKT2020}. \\

Cicoli et al. \cite{Cicoli20} have shown the existence of new accelerating solution in late time cosmology even for steep potentials. Using the technique of dynamical system analysis, Chakraborty et al. \cite{Chakraborty20} have analyzed the cosmological inference without solving the coupled cosmic evolution equation. Using the analytical degeneracy relation between cosmological parameters and numerical fits to the cosmological data Alestas et al. \cite{Alestas20} have identified the quantitative and qualitative features of the dark energy models. The dark energy model by Dantas and Rodrigues \cite{Dantas20} have highlighted the common parametrizations for the equation of state as a function of redshift in the context of twin-like theories. Cheng et al. \cite{Cheng20} have shown the possible interaction between dark energy and dark matter and its possible impact on the cosmic evolution of the Universe. Paliathansis and Leon \cite{Paliathanasis20} have determined the equilibrium points for the field equations of Bianchi I space-time in Einstein-aether scalar field theory and studied the dynamics and evolution of anisotropies. Lin and Qian \cite{Lin20} presented a dark energy model and have shown that the eventual fate of the Universe is largely insensitive to the initial conditions, and hence the cosmic coincidence problem can be avoided.\\

Padmanabhan and Chitre \cite{Padmanabhan87} have indicated the prime role of viscosity at late epoch in the accelerated expansion of the Universe. Zimdahl et al. \cite{Zimdahl01} have indicated that cosmic anti-friction leads to a negative bulk pressure, which may be accounted for the magnitude-redshift data of type Ia supernovae. Cataldo et al. \cite{Cataldo05} have shown that the dark energy of the Universe to be phantom energy due to the negative pressure generated by the bulk viscosity. Brevik and Gorbunova \cite{Brevik05} have shown that with the addition of large bulk viscosity, the fluid which lies in the quintessence region ($\omega_{de}>-1 $) reduces the thermodynamical pressure and behaves like a Phantom fluid ($\omega_{de}<-1 $). Brevik et al. \cite{Brevik11} have proved that the viscous fluid is able to produce a little rip cosmology purely as the viscous effect whereas with turbulence approach, Brevik et al. \cite{Brevik12} have shown the occurrence of both big rip and little rip behavior of dark energy. Velten et al. \cite{Velten13} have shown the effect of bulk viscosity for the phantom dark energy. Kumar \cite{Kumar13} has studied the dynamical behaviour of the universe in spatially homogeneous space-time with non-interacting matter and dark energy component. Amirhashchi \cite{Amirhashchi14} has considered the anisotropic space-time and studied the interaction between dark matter dark energy and dark matter in the scope of viscous matter field. Rezaei and Malekjani \cite{Rezaei17} have studied agegraphic dark energy models and have shown the affect of dark energy in the growth of large scale structures of the Universe. Wang et al. \cite{Wang17} have constrained the viscous dark energy models based on the observational data. Amirhashchi \cite{Amirhashchi17} has mapped the Bianchi V space time to FRW metric and has shown how the bulk viscosity affect the dark energy equation of state (EoS) parameter. Atreya et al. \cite{Atreya19} have studied the viscus dark energy model in a small range of the redshift $0\leq z\leq 2.5$.  Mishra et al. \cite{Mishra19b} have developed a general form of skewness parameters and have shown the dominance of viscous fluid and dark energy fluid respectively in early time and late time of evolution. Silva and Silva \cite{Silva19} have proposed the non-extensive effects of Verlinde theory to construct an extended cosmological model with viscous dark energy. Odintsov et al. \cite{Odintsov20} have shown that under the negative value of bulk viscosity, the power law model does not exhibit good results whereas the logarithmic model provides a good fit to the $\Lambda$CDM model. \\

Hinshaw et al. \cite{Hinshaw03,Hinshaw07} have suggested that the metric components be of different functions of time because there is a small variation observed between the intensities of the microwaves received from different directions. In order to compare the detail observations, space-time that behaves almost like FRW metric needs to be considered. Bianchi V space-time which are spatially anisotropic but homogeneous may be more suitable for this purpose.  Therefore, we consider here Bianchi type V (BV) space time in the form

\begin{equation} \label{eq1}
ds^{2}= dt^{2}- A^{2}dx^{2}- e^{2 \alpha x} \left(B^2dy^{2}+C^{2}dz^{2}\right)
\end{equation} 
where $A=A(t)$, $B=B(t)$ and $C=C(t)$. The exponent $\alpha (\neq0)$ is an arbitrary constant. For $\alpha = 0$, the space-time \eqref{eq1} reduces to  Bianchi I metric.

We organize the present work as follows. In section II, the field equations in non-interacting dark energy and bulk viscous fluid are derived. In section III, a model with viscous matter energy density has been formulated and the observational constraints on the model parameters are presented in section IV. The physical behavior of the model with the observational data such as age of the Universe, deceleration parameter, particle horizon, jerk parameter, $om(z)$ parameter are presented and analysed in section V.  Results and discussions are given in section VI.
\section{Basic Governing Equations}      

We assume that our Universe be filled with viscous matter field and dark energy fluid so that the total energy momentum tensor in the two fluid scenario can be expressed as

\begin{equation}
T_{ij}= T_{ij}^{m} + T_{ij}^{de}
\end{equation}
where, $T_{ij}^{m}$ and $T_{ij}^{de}$ respectively denote the energy momentum tensor of barotropic viscous fluid and dark energy fluid and can be expressed as

\begin{eqnarray}
T_{ij}^{m} &=& (\rho+\bar{p})u_iu_j-\bar{p} g_{ij}\nonumber\\
           &=& (\rho + {p-3\zeta H^2 }) u_{i}u_{j}- {(p- 3\zeta H^2) } g_{ij},\\
T_{ij}^{de} &=& diag[\rho^{de}, -p^{de}, -p^{de}, -p^{de}]
\end{eqnarray}
where, the effective pressure $\bar{p}=p-3\zeta H^2$ consists of  proper pressure and  barotropic bulk viscous pressure. $3\zeta H^2$ is the bulk viscous pressure.  While $H$ represents the Hubble rate, $\zeta$ is the coefficient of bulk viscosity. Here, $u^{i}$ is the four velocity vector of the fluid in a co-moving coordinate system. $\rho$ and $p$ respectively be the energy density and pressure of the matter field whereas $\rho_{de}$ and $p_{de}$ respectively represent the energy density and pressure of dark energy. Now, the Einstein's field equations $G_{ij}=-8\pi T_{ij}$ for the two fluid scenario for the metric \eqref{eq1} can be obtained as, 
\begin{eqnarray}
\frac{\ddot{B}}{B}+\frac{\ddot{C}}{C}+\frac{\dot{B}\dot{C}}{BC}-\frac{\alpha^{2}}{A^{2}}&=& -8\pi G(p -3\zeta H^2 +p_{de}) \label{fe2} \\
\frac{\ddot{A}}{A}+\frac{\ddot{C}}{C}+\frac{\dot{A}\dot{C}}{AC}-\frac{\alpha^{2}}{A^{2}}&=& -8\pi G(p -3\zeta H^2+p_{de})\label{fe3} \\
\frac{\ddot{A}}{A}+\frac{\ddot{B}}{B}+\frac{\dot{A}\dot{B}}{AB}-\frac{\alpha^{2}}{A^{2}}&=&-8\pi G(p -3\zeta H^2+p_{de}) \label{fe4} \\
\frac{\dot{A}\dot{B}}{AB}+\frac{\dot{B}\dot{C}}{BC}+\frac{\dot{C}\dot{A}}{CA}-\frac{3\alpha^{2}}{A^{2}}&=& 8\pi G(\rho+\rho_{de}) \label{fe5}\\
2\dfrac{\dot{A}}{A}-\dfrac{\dot{B}}{B}-\dfrac{\dot{C}}{C}&=&0.  \label{fe6}
\end{eqnarray}
Here an over dot represents the derivative of corresponding field variable with respect to $t$. 
Eq.\eqref{fe6} implies $A^2=BC$, where the proportionality constant has been absorbed into the metric potential.

With an algebraic manipulations among eqs. (\ref{fe2})-(\ref{fe6}), we obtained the following relations among the metric potentials,\\
\begin{center}
$B = A\varepsilon,\;\; C = \frac{A}{\varepsilon},\;\;\;\;\frac{\dot{\varepsilon}}{\varepsilon} =\frac{k}{A^3}$
\end{center}
where $\varepsilon=\varepsilon(t)$ and it measures the anisotropy of the Universe. $k$ is an integration constant. If we take $\varepsilon = 1$ then $A$ = $B$ = $C$ and the Universe will expand with uniform rate in $x$, $y$ and $z$ directions. We can now recast eqs. \eqref{fe2}-\eqref{fe5} as
\begin{eqnarray}
2\frac{\ddot{A}}{A}+\frac{\dot{A^{2}}}{A^2} &=& -8\pi G(p_m -3\zeta H^2 +p_{de}+p_{\sigma}+p_{\alpha})\label{fe7} \\
H^2 = \frac{\dot{A^{2}}}{A^2} &=& \frac{8\pi G}{3}(\rho_m+\rho_{de}+\rho_{\sigma}+\rho_{\alpha}) \label{fe8}
\end{eqnarray}
Here, we have made certain assumptions such as\\
\begin{center}
$ p_{\sigma}=\rho_{\sigma}=\frac{k^{2}}{8\pi GA^{6}},~~~ p_{\alpha} =  -\frac{{\alpha}^2}{8 \pi G A^2},~~~ \rho_{\alpha}= \frac{3{\alpha}^2}{8 \pi G A^2}$.
 \end{center}

The energy conservation equation can be expressed as
\begin{equation} \label{ece}
T_{;j}^{ij}=\dot{\rho}+3H(p+\rho)=0
\end{equation}
where, $\rho=\rho_{m}+\rho_{de}+\rho_{\sigma}+\rho_{\alpha}$ and $p=p_{m}-3\zeta H^2 +p_{de}+p_{\sigma}+p_{\alpha}$. It is noteworthy to mention here that, the energy conservation equations due to anisotropy of Universe and $\alpha$ parameter are hold separately as

\begin{eqnarray} \nonumber
\dot{\rho}_{\sigma}+3H(p_{\sigma}+\rho_{\sigma})=0 \\
\dot{\rho}_{\alpha}+3H(p_{\alpha}+\rho_{\alpha})=0 \nonumber
\end{eqnarray}

Finally, we assumed the non-interacting scenario between dark energy and viscous fluid, hence the continuity equations $\frac{d}{dt}{(\rho_{m}+\rho_{de})}+3H(p_{m}+p_{de}+\rho_{m}+\rho_{de})=0$ due to viscous fluid  and dark energy are conserved separately as 
\begin{eqnarray}
\dot{\rho}_{m}+3H(p_{m}-3\zeta H^2+\rho_{m})=0 \\
\dot{\rho}_{de}+3H(p_{de}+\rho_{de})=0
\end{eqnarray}

Since the present Universe is almost filled with pressureless  matter (dust) for which the equation of state (EoS) parameter, $\omega_{m}=0$, the continuity equation due to viscous matter fluid can be reduced to  $\dot{\rho}_{m}+3H(\rho_{m})=3\zeta H^3$.

\section{Viscous matter energy}
Now we use the relationship between the scale factor $a$ and red shift $z$ as, $\frac{a_0}{a} = 1+z$ which enables us to obtain 

\begin{eqnarray}
\rho_{\sigma}&=&(\rho_{\sigma})_0 \left[\frac{a_0}{a}\right]^6 = (\rho_{\sigma})_0 (1+z)^6, \\
\rho_{de}&=&(\rho_{de})_0\left[\frac{a_0}{a}\right]^{3(1+\omega_{de})}=(\rho_{de})_0 (1+z)^{3(1+\omega_{de}}),
\end{eqnarray}
where $\omega_{de}$ is the EoS parameter for dark energy which is considered as constant for present epoch.
We take $p_m=0$ for dust filled universe and define the following energy parameters~

\begin{center}
$\Omega_{m}=\frac{\rho_{m}}{\rho_{c}}$,~~ $\Omega_{de}=\frac{\rho_{de}}{\rho_{c}}$,~~
$\Omega_{\sigma}=\frac{\rho_{\sigma}}{\rho_{c}}$,
\end{center}
where, $\rho_{c}=\frac{3H^{2}}{8\pi G}$, $p_{\sigma}=\rho_{\sigma}=\frac{k^{2}}{8\pi Ga^{6}}$, $p_{\alpha}= -\frac{1}{3}  \rho_{\alpha}=-\frac{\alpha^{2}}{8\pi Ga^{2}}$,
$\rho_{\alpha}= (\rho_{\alpha})_0  [\frac{a_0}{a}]^{2}= (\rho_{\alpha})_0
(1+z)^2$, $\rho_{\sigma}= (\rho_{\sigma})_0  [\frac{a_0}{a}]^6= (\rho_{\sigma})_0
(1+z)^6$. 

Now, from eq. \eqref{fe8}, we can obtain the energy relation and the differential equation for the Hubble parameter 
$H$ respectively as,
\begin{equation}\label{eq17}
(\Omega_{m})+(\Omega_{de})+(\Omega_{\sigma}) + (\Omega_{\alpha}) = 1
\end{equation}
and
\begin{equation} \label{eq18}
3(8\pi\zeta-1)H^2+(z+1)(H^2)'= 3H_0^2\left(\omega_{de}(\Omega_{de})_0(1+z)^{3(1+\omega_{de})}+\omega_{\sigma}(\Omega_{\sigma})_0(1+z)^{3(1+\omega_{\sigma})}+\omega_{\alpha}(\Omega_{\alpha})_0(1+z)^{3(1+\omega_{\alpha})}\right)
\end{equation}
where prime ($\prime$) denotes differentiation with respect to redshift $z$.\\ 
Thus,  eq. \eqref{eq18} can be rewritten as
\begin{equation}\label{fe9}
3(8\pi\zeta-1)H^2+(z+1)(H^2)'= 3H_0^2\left(\omega_{de}(\Omega_{de})_0(1+z)^{3(1+\omega_{de})}+(\Omega_{\sigma})_0(1+z)^{6}-\frac{1}{3}(\Omega_{\alpha})_0(1+z)^{2}\right)
\end{equation}
On solving (\ref{fe9}), we obtain the following expression for Hubble parameter $H(z)$ 
\begin{eqnarray}\label{fe10}
H(z) &=& H_0\sqrt{\left(\frac{\omega_{de}(\Omega_{de})_0 (z+1)^{3(1+\omega_{de})}} {8\pi\zeta+\omega_{de}}+\frac{\omega _{\sigma} (\Omega_{\sigma})_0(z+1)^{3(1+\omega_{\sigma})}}{8\pi\zeta+\omega_{\sigma}} +\frac{\omega _{\alpha}(\Omega_{\alpha} )_0(z+1)^{3(1+\omega_{\alpha})}}{8\pi\zeta+\omega_{\alpha}}\right)} \nonumber \\ 
~~~~~~~~~&+&\overline{\left(1 - \frac{\omega_{de}(\Omega_{de})_0}{\omega_{de}+8\pi \zeta}
-\frac{\omega_{\sigma}(\Omega_{\sigma})_0}{\omega_{\sigma}+8\pi \zeta}
-\frac{\omega_{\alpha}(\Omega_{\alpha})_0}{\omega_{\alpha}+8\pi\zeta}\right) (z+1)^{3(1-8\pi\zeta)}}
\end{eqnarray}
The energy parameter ($\Omega_m$) and the energy density ($\rho_m$) are read as
\begin{equation*}\label{key}
\begin{split}
\Omega_{m} & =(\Omega_{m})_0 \left(-8\pi\zeta\left(\frac{ (z+1)^{3(1+\omega_{de})}} {8\pi\zeta+\omega_{de}}+\frac{(z+1)^{3(1+\omega_{\sigma})}}{8\pi\zeta+\omega_{\sigma}} +\frac{(z+1)^{3(1+\omega_{\alpha})}}{8\pi\zeta+\omega_{\alpha}}\right) \right. \\
&\quad\quad\quad \left. {} +\left(1 - \frac{\omega_{de}(\Omega_{de})_0}{\omega_{de}+8\pi \zeta}
	-\frac{\omega_{\sigma}(\Omega_{\sigma})_0}{\omega_{\sigma}+8\pi \zeta}
	-\frac{\omega_{\alpha}(\Omega_{\alpha})_0}{\omega_{\alpha}+8\pi\zeta}\right) (z+1)^{3(1-8\pi\zeta)}\right)
\end{split}
\end{equation*}
\begin{equation}
\begin{split}
&\times\left( \frac{\omega_{de}(\Omega_{de})_0 (z+1)^{3(1+\omega_{de})}} {8\pi\zeta+\omega_{de}}+\frac{\omega _{\sigma} (\Omega_{\sigma})_0(z+1)^{3(1+\omega_{\sigma})}}{8\pi\zeta+\omega_{\sigma}} +\frac{\omega _{\alpha} (\Omega_{\alpha} )_0(z+1)^{3(1+\omega_{\alpha})}}{8\pi\zeta+\omega_{\alpha}} \right. \\
&\quad\quad\quad \left. {} + \left(1 - \frac{\omega_{de}(\Omega_{de})_0}{\omega_{de}+8\pi \zeta}
-\frac{\omega_{\sigma}(\Omega_{\sigma})_0}{\omega_{\sigma}+8\pi \zeta}
-\frac{\omega_{\alpha}(\Omega_{\alpha})_0}{\omega_{\alpha}+8\pi\zeta}\right) (z+1)^{3(1-8\pi\zeta)}\right)^{-1}
\end{split}
\end{equation}

\begin{eqnarray}
\rho_{m}& =&(\rho_{m})_0 \times(-8\pi\zeta)\left(\frac{ (z+1)^{3(1+\omega_{de})}} {8\pi\zeta+\omega_{de}}+\frac{(z+1)^{3(1+\omega_{\sigma})}}{8\pi\zeta+\omega_{\sigma}} +\frac{(z+1)^{3(1+\omega_{\alpha})}}{8\pi\zeta+\omega_{\alpha}}\right) \nonumber \\
&+&(\rho_{m})_0\times\left(1 - \frac{\omega_{de}(\Omega_{de})_0}{\omega_{de}+8\pi \zeta}
-\frac{\omega_{\sigma}(\Omega_{\sigma})_0}{\omega_{\sigma}+8\pi \zeta}
-\frac{\omega_{\alpha}(\Omega_{\alpha})_0}{\omega_{\alpha}+8\pi\zeta}\right) (z+1)^{3(1-8\pi\zeta)}
\end{eqnarray}

From eq. \eqref{fe7}, we obtain

\begin{equation}\label{fe11}
(2q-1) H^2 = 3 H^2_0 \left(\omega_{de}(\Omega_{de})_0(1+z)^{3(1+\omega_{de})} + (\Omega_{\sigma})_0 (1+z)^6 -\frac{1}{3} (\Omega_{\alpha})_0 (1+z)^2  \right)
\end{equation}
where, $q = -\frac{\ddot{a}}{aH^2}$ is the deceleration parameter.\\

\section{Observational constraints on model parameter of viscous dark energy universe}
In this section, we describe $H(z)$ and Pantheon observational data and the statistical methodological analysis for constraining the viscous dark energy Universe .
\begin{itemize}
\item {\bf Observational Hubble Data (OHD)}: We have taken over $46~H(z)$ observational data points in the range of $0\leq z\leq 2.36$,  dominated from cosmic chronometric technique. These all $46~H(z)$ data points are compiled in Table I of Ref. \cite{Biswas/2019}.\\

\item {\bf Pantheon data}: We use the Pantheon compilation \cite{Scolnic/2018} which includes 1048 SNIa apparent magnitude measurements including 276 SNIa $(0.03 < z < 0.65)$ investigated by the Pan-STARRS1 Medium Deep Survey and SNIa distance estimates from SDSS, SNLS and low-z HST samples.\\
\end{itemize}
To estimating the free parameters of the model by bounding it with observational data points, we define $\chi^{2}$ as following:
\begin{equation}
\label{chi1}
\chi^{2} = \sum_{i=1}^{N}\left[\frac{E_{th}(z_{i})-E_{obs}(z_{i})}{\sigma_{i}}\right]^{2}
\end{equation}
where $E_{obs}(z_{i})$ \& $E_{th}(z_{i})$ are the observed values and the theoretical values of corresponding parameters with errors $\sigma_{i}$. N denotes total number of data points.\\

For joint analysis, $\chi^{2}_{joint}$ can read as
\begin{equation}
\label{joint}
\chi^{2}_{joint} = \chi^{2}_{H(z)} + \chi^{2}_{Pantheon}
\end{equation}
 
Finally, in our analysis, we assume the following uniform priors for free parameters\\
\begin{equation}
\label{priors}
H_{0}\sim U(50,\; 80),\;\; (\Omega_{de})_{0} \sim U(0.5,\; 0.8)\; \&\;\; \omega_{de} \sim U (-0.5, -1.5)
\end{equation}
\begin{figure}[ht!]
\centering
\includegraphics[width=6cm,height=6cm,angle=0]{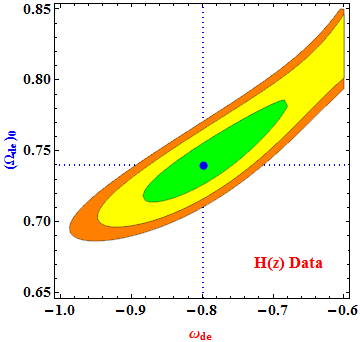} 
\caption{Two dimensional contours at $1\sigma$, $2\sigma $ and  $3\sigma$ confidence levels by bounding our model with latest 46 $H(z)$ data. The estimated parameters are $H_0=69.39 \pm 1.54~km~s^{-1}Mpc^{-1}$, $(\Omega_{de})_{0} = 0.74 \pm 0.04$ and $\omega_{de} = -0.799 \pm 0.08$.}
\end{figure}
\begin{figure}[ht!]
\centering
\includegraphics[width=6cm,height=6cm,angle=0]{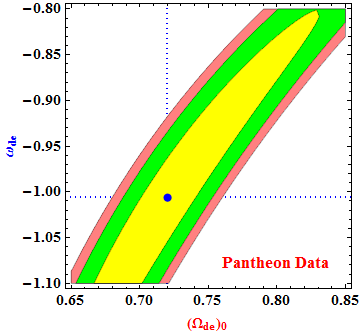}  
\caption{Two dimensional contours at $1\sigma$, $2\sigma $ and  $3\sigma$ confidence levels by bounding our model with  Pantheon data. The estimated parameters are $H_{0} = 70.016 \pm 1.65;km\;s^{-1}\;Mpc^{-1}$, $(\Omega_{de})_{0} = 0.72 \pm 0.05$ and $\omega_{de} = -1.006 \pm 0.04$.}
\end{figure}
\begin{figure}[ht!]
\centering
\includegraphics[width=6cm,height=6cm,angle=0]{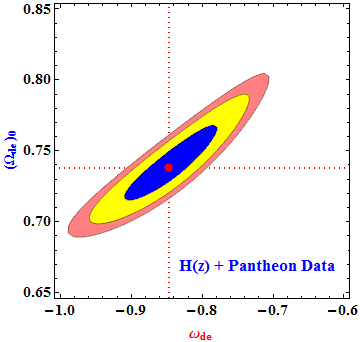} 
\caption{Two dimensional contours at $1\sigma$, $2\sigma $ and  $3\sigma$ confidence levels by bounding our model with  joint $H(z)$ and Pantheon data. The estimated parameters are $H_{0} = 69.36 \pm 1.42~km\;s^{-1}\;Mpc^{-1}$, $(\Omega_{de})_{0} = 0.738 \pm 0.042$ and $\omega_{de} = -0.847 \pm 0.043$.}
\end{figure}
The summary of numerical analysis is listed in Table 1. The two dimensional contours at $1\sigma$, $2\sigma $ and  $3\sigma$ confidence regions by bounding our model with latest 46 $H(z)$ data, Pantheon data and joint $H(z)$ and Pantheon data are depicted in Fig. 1, Fig. 2 and Fig. 3 respectively. It is interesting to mention here that, within $1\sigma$ error our analysis provides good estimates of the dark energy density parameter as $0.74\pm 0.04$, $0.72\pm 0.05$ and $0.738\pm 0.042$ from the $H(z)$, Pantheon data and a joint analysis respectively. These data sets also predict the EoS parameter for our derived model respectively as $-0.799\pm 0.08$, $-1.006\pm 0.04$ and $-0.847\pm 0.043$. One may note that, the predictions from the Pantheon comoilation data set from our derived model are closer to the concordance $\Lambda$CDM model. The best fit curve of Hubble rate versus redshift $z$ with $46$ observational Hubble data of derived model is shown in Fig. 4. It may be noticed from the figure that, our model passes almost in the middle of the observational $H(z)$ data points. \\

So far the $\Lambda$CDM model provides a good description of the observable Universe and is considered as a successful model. However, it suffers from the well known $H_0$ tension. The $H_0$ tension arises because of the discrepancy between the local distance ladder measurement of the Hubble Space Telescope (HST) observations of 70 long-periods Cepheids in the large Magellanic Cloud \cite{Riess2016, Riess2018, Riess2019} and an indirect measurement based on $\Lambda$CDM and the CMB temperature from Planck collaboration \cite{Planck2020}. While the measurement of Riess  et al. yields   $H_0 = 74.03\pm 1.42~km~s^{-1}Mpc^{-1}$  \cite{Riess2016}, the Planck collaboration data provides $H_0=67.36\pm 0.54 ~km~s^{-1}Mpc^{-1}$ \cite{Planck2020}. The statistical significance of the $H_0$ tension is $4.4\sigma$. Other estimates of the Hubble parameter include that from S$H0$ES, $H_0=73.5\pm 1.4~km~s^{-1}Mpc^{-1}$ \cite{Reid2019} and that from $H0$LiCOW collaboration, $H_0=73.3\pm 1.7~km~s^{-1}Mpc^{-1}$ \cite{Wong2019}. The Hubble parameter as measured from the gravitational wave event GW170817 and its electromagnetic counterpart GRB170817A is $H_0=70.0^{+12.0}_{-8.0}~km~s^{-1}Mpc^{-1}$ \cite{LIGO2017}. Basing upon a calibration on the  Tip of the Red Giant Branch (TRGB), applied to SNIa, Freedman et al. obtained a value $H_0=69.8 \pm 0.8 (\pm 1.1\%~ \text{stat})\pm 1.7(\pm 2.4\%~ \text{sys})~km~s^{-1}Mpc^{-1}$ \cite{Freedman2019}. In recent years there have been many attempts to reduce the $6~km~s^{-1}Mpc^{-1}$ tension using different mechanisms \cite{Lambiase2019, Sola2020, Adhikari2020}.  In the context of a possible explanation to the discrepancy, two obvious questions are raised: (a) whether or not we live in an underdense local void or bubble and (b) the effects of weak lensing on the SNe Ia measurements. It is believed that, the $H_0$ tension may hint for a new Physics involving the dark energy and/or dark matter components either beyond the standard $\Lambda$CDM model, or beyond the standard model of particle physics \cite{Bernal2016, Mortsell2018, Freedman2019}. In this work, we have estimated the Hubble parameter at the present epoch as $H_0=69.39 \pm 1.54~km~s^{-1}Mpc^{-1}$, $H_0=70.016 \pm 1.65~km~s^{-1}Mpc^{-1}$ and $H_0=69.36 \pm 1.42~km~s^{-1}Mpc^{-1}$ respectively from the $H(z)$ data, Pantheon data and the combined analysis of $H(z)$+Pantheon data. The estimated ranges of the Hubble parameter from our analysis are well within the recent estimates mentioned earlier and lie in the middle of the range of the Hubble tension.

\begin{table*}
\small
\caption{Summary of the numerical result. \label{tbl-1}}
\begin{tabular}{@{}lrrrrrrrrrrr@{}}
\hline
Source/Data ~~~~ & ~~~~ $H(z=0)$ ~~~~~ & ~~~~~ Pantheon ~~~~~ & ~~~~~ $H(z)$ + Pantheon \\
\hline
$H_{0} (~km~s^{-1}Mpc^{-1}) $ ~~~~ &~~~~ $69.39 \pm 1.54$ ~~~~ & ~~~~ $70.016 \pm 1.65$ ~~~~ & ~~~~ $69.36 \pm 1.42$\\
$(\Omega_{de})_{0}$ ~~~~ &~~~~ $0.74 \pm 0.04$ ~~~~ & ~~~~ $0.72 \pm 0.05$ ~~~~ & ~~~~ $0.738 \pm 0.042$\\
$\omega_{de}$ ~~~~ &~~~~ $-0.799 \pm 0.08$ ~~~~ & ~~~~ $-1.006 \pm 0.04$ ~~~~ & ~~~~ $-0.847 \pm 0.043$\\ \\
\hline
\end{tabular}
\end{table*}  
\begin{figure}[t!]
\centering
\includegraphics[width=8cm,height=7cm,angle=0]{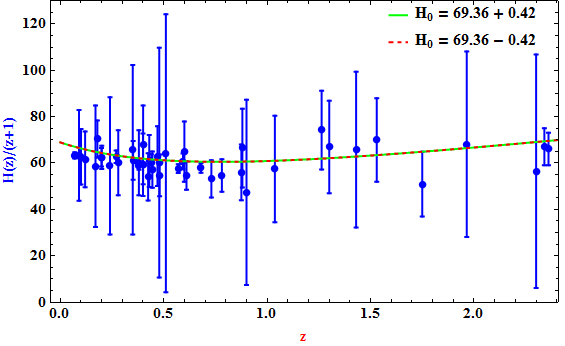} 
\caption{The plot of Hubble rate $H(z)/(1+z)$ versus $z$ for $H_{0} = 69.36 \pm 0.42\;km\;s^{-1}\;Mpc^{-1}$, $(\Omega_{de})_{0} = 0.738$ and $\omega_{de} = -0.847$. The points with error bars indicate the observed Hubble data. The solid green  and the dotted red line represent our derived model.}
\end{figure}
\section{Physical behaviour of model}
\subsection{Age of the Universe}
The age of the universe is computed as
$$H_0 (t_0-t) = \int_{0}^{z}\frac{dx}{(1+x) h(x)},~ h(z)=H(z)/H_{0}$$\\
where, $$H_0 t_0 =\lim_{z\rightarrow\infty}\int_{0}^{z}\frac{dx}{(1+x) h(x)}.$$
$H(z)$ is given in eq. (\ref{fe10}).\\
Fig. 5 displays the behaviour of time with red shift. It is obtained that $H_0 (t_0-t) $ converges to 0.9796 for infinitely large $z$. We translate this to find out the present age of the Universe as $t_0= 0.9796H_0^{-1}\sim 13.79$ Gyrs which is quite close (within $0.03\%$) to the age of the Universe estimated from Planck results $t_0=13.786\pm 0.020$ Gyrs \cite{Planck2020}. Therefore, the derived model of the Universe has pretty consistency with recent observations. We also note that earlier Planck results predicts present age of the Universe as $13.81\pm 0.038$ Gyrs \cite{Ade16}. In some recent investigations, the present age of the Universe are also reported in the vicinity of estimated age of the Universe in this paper (see Refs. \cite{Yadav/2021,Prasad/2020Pramana,Prasad/2020}). 
\begin{figure}[ht!]
\centering
\includegraphics[width=7cm,height=6cm,angle=0]{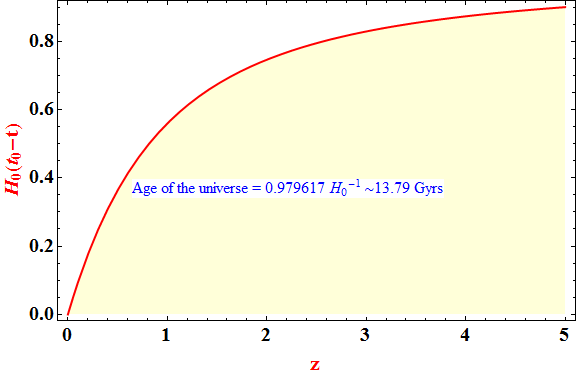} 
\caption{The plot of time $t$ over red-shift $z$.}
\end{figure}

\subsection{Deceleration parameter}
From Eq. (\ref{fe11}), we obtain the expression for deceleration parameter $q$ as following
\begin{equation}\label{dec}
q(z)=\frac{1.5 H^2_0 \left(\omega_{de} (\Omega_{de})_0 (z+1)^{\{3 (\omega_{de}+1)\}}-0.333(\Omega_{\alpha})_0  (z+1)^2+(\Omega_{\sigma})_0  (z+1)^6\right)}{H(z)^2}+0.5
\end{equation}
Fig. 6 describes the dynamics of deceleration parameter $q$ with respect to red-shift $z$.
\begin{figure}[ht!]
\centering
\includegraphics[width=8cm,height=6 cm,angle=0]{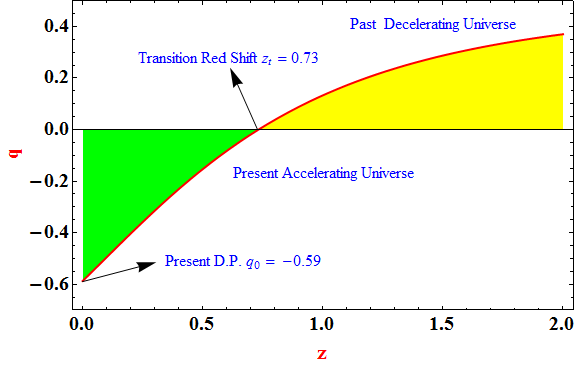} 
\caption{The plot of deceleration parameter $q$ over red shift $z$.}
\end{figure}
The deceleration parameter evolves from a positive domain to a negative domain. The transition from a decelerated phase to an accelerated phase occurs at the transition redshift $z_t=0.73$.  This transition redshift value is compatible to the recently constrained value  $z_t=0.74\pm 0.05$ of Farooq and Ratra \cite{Farooq2013}, $z_t=0.82\pm 0.08$ of Busca \cite{Busca2013}, $z_t=0.69^{+0.23}_{-0.12}$ of Lu et al. \cite{Lu2011}, $z_t=0.60^{+0.21}_{-0.12}$ of Yang and Gong \cite{Yang2019} and  $z_t=0.7679^{+0.1831}_{-0.1829}$ of Capozziello et al \cite{Capo2014}. In a recent analysis, the deceleration parameter at the present epoch is obtained to be $q_0=-1.08\pm 0.29$ \cite{Camarena2020}. Using our anisotropic model, the present value of deceleration parameter is computed as $q_0=-0.59$ which is  quite closer to the values constrained in a recent work \cite{Capo2020}. The transit redshift $z_t=0.73$ can be translated into cosmic time through the time red-shift relation as described in above subsection.  In fact, from our present analysis, we infer that, the deceleration to acceleration has occurred around 5.9 Gyrs from now.\\ 
\subsection{Particle horizon}
Particle Horizon is defined as the maximum distance from which light from particles have travelled to the observer and defines the size of the observable Universe. Following Bentabol et al \cite{Bentabol/2013}, the particle horizon can be  obtained as
\begin{equation}
\label{ph}
R_{p} = lim_{t_{p}\rightarrow 0}\;\;a_{0}\int_{t_{p}}^{t_{0}}\frac{dt}{a(t)} = lim_{z\rightarrow \infty}\int_{0}^{z}\frac{dz}{H(z)}
\end{equation}
where $t_{p}$ denotes time in past at which the light signal was emitted from source.\\
A comprehensive analysis of particle horizon for anisotropic Universe is given in Refs. \cite{Goswami/2020,Prasad/2020}.\\
\begin{figure}[h!]
\centering
\includegraphics[width=7cm,height=6cm,angle=0]{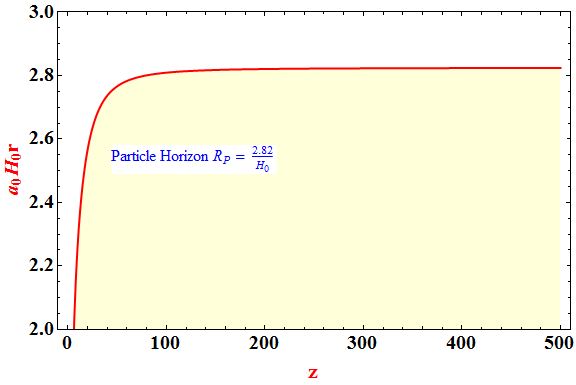}
\caption{The plot of Proper distance $a_{0}~H_{0}~r$ versus $z$.}
\end{figure}
Integrating equation (\ref{ph}), we obtain
\begin{equation}
\label{ph-2}
R_{p} = \frac{2.82455}{H_{0}}
\end{equation}
Thus, for derived model, the particle horizon $R_{p} = \frac{2.82455}{H_{0}}$. This means that at present, we are able to see only those galaxies whose proper distances from us happen to less than $\frac{2.82455}{H_{0}}$. The variation of $a_{0} H_{0} r$ versus $z$ is shown in Fig. 7. Here $r$ measures the proper distance. 
\subsection{Jerk parameter} 
The dimensionless jerk parameter $j=\frac{\dddot{a}}{aH^2}$ involves the third order derivative of the scale factor and provides a suitable kinematic approach to test cosmological models. It is worth to mention here that the kinematic approaches are advantageous because of their model independence in the sense of cosmic composition. Also, it helps to search for a possible deviation from the $\Lambda$CDM model. The $\Lambda$CDM model favours $j_0=1$ at the present epoch. For $j=1$, it is meant that, the universe continues to have accelerated expansion under a cosmological constant.  Any deviation from this value obviously implies a non-$\Lambda$CDM model \cite{Sahni2003, Alam2003}. In terms of the deceleration parameter, $j$ is defined as
\begin{equation}
j(q,z)=(2q+1)q+(1+z)q^{\prime},
\end{equation}
where $q^{\prime}$ is the derivative with respect to the redshift. In terms of Hubble parameter, the jerk parameter can be obtained as
\begin{equation}
j(z)=\frac{1}{2}(1+z)^2\frac{ \left[H(z)^2\right]^{\prime\prime}}{H(z)^2}-(1+z)\frac{\left[H(z)^2\right]^{\prime}}{H(z)}+1,
\end{equation}
where the prime denotes the derivative with respect to redshift. The value of the jerk parameter at the present epoch may be read as $j_0=(2q_0+1)q_0+q^{\prime}(z=0)$, where $q_0$ is the deceleration parameter at the present epoch. Since the jerk parameter appears as an important quantity in the kinematic approach, many authors have used different parametrizations in literature and constrained the constants appearing in the parametrization from observational data such the $H(z)$ data, type Ia Supernova or CMB or BAO dataset\cite{Zhai2013, Mamon2018, Mukherjee2017}. However, in the present work, we have used the $H(z)$ and Pantheon data set to constrain the jerk parameter without assuming any specific parametrization  in terms of redshift. Such an approach provides an unbiased constraint on the jerk parameter. In Figure 8, we have shown the evolution of the jerk parameter $j(z)$ within $1\sigma$ error for $\xi = 0.002$ and $\alpha = 0.0001$.\\

In each panel, the central dark line denotes the best fit curve. We observe that, the jerk parameter evolves in a domain of positive values. From the reconstructed plots of the jerk parameter, it is inferred that, the extracted values of $j(z)$ within $1\sigma$ error deviate from the concordance $\Lambda$CDM value $j_{0} = 1$ at $z=0$.  However, the extracted value of $j$ from the Pantheon data set is bit closer to the  $\Lambda$CDM value compared to that constrained from the $H(z)$ data and the combined analysis of $H((z))$ and Pantheon data. In particular we obtained $j_0 = 0.384\pm 0.112$, $j_0 = 0.968\pm 0.129$  and $j_0 = 0.496\pm 0.116$ for the $H(z)$, Pantheon and $H(z)$ + Pantheon data respectively. Since, the late time cosmic speed up  issue has yet to be resolved completely with the cause and origin is still speculative, the deviation of the jerk parameter for the present anisotropic model from the $\Lambda$CDM value  requires an involved attention. The value of the jerk parameter at the present epoch should be obtained from the current or some low-redshift observations and since the low-redshift observations are not comprehensive, the exact determination of $j$ value at the present epoch is pushed into bare uncertainty \cite{Zhai2013}. Mamon and Bamba used the $H(z)$ observational data and constrained the jerk parameter by considering some parametrized form of the deceleration parameter \cite{Mamon2018}. They have also observed a departure of the jerk parameter value from the concordance  $\Lambda$CDM value. Capozziello et al. \cite{Capo2020} have also obtained a slightly larger value of $j$ than the value predicted by concordance paradigm. Mukherjee and Banerjee in their work \cite{Mukherjee2017}, concluded that, any deviation of the jerk parameter from the $\Lambda$CDM value indicates an interaction between the dark energy and dark matter.\\ 
\begin{figure}[t!]
\centering
\includegraphics[width=5cm,height=4cm,angle=0]{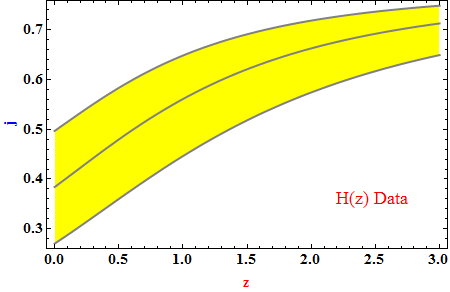}
\includegraphics[width=5cm,height=4cm,angle=0]{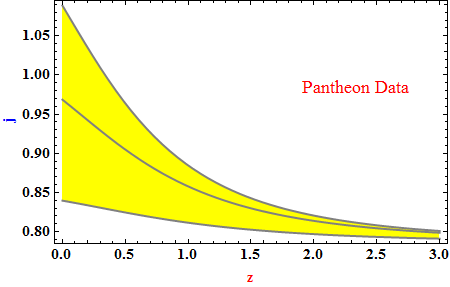}
\includegraphics[width=5cm,height=4cm,angle=0]{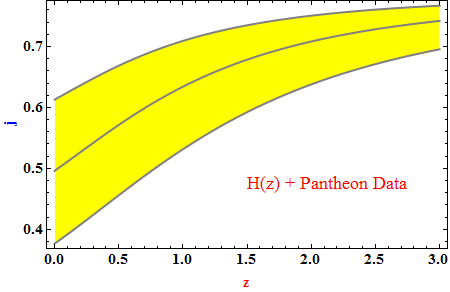} 
\caption{The plot of jerk parameter $j$ as function of red-shift $z$ are shown in $1\sigma$ error for $\xi = 0.002$ and $\alpha = 0.0001$. In each panel, central dark line denotes the best fit curve.}
\end{figure}
\subsection{Om(z) parameter} 
$Om(z)$ parameter is considered as another powerful diagnostic tool for different dark energy models and is defined as \cite{Sahni2008,Sahni2014}, 
\begin{equation}
Om(z)=\frac{H(z)^2-H_{0}^2}{H_{0}^2 \left[(z+1)^3-1\right]}.
\end{equation}
The excellent features of the $Om$ diagnostic is that, it effectively distinguishes the dynamical dark energy models from $\Lambda$CDM model and ensures $Om(z)=\Omega_{m0}$ for a null test hypothesis of $\Lambda$CDM model. For a constant dark energy EoS parameter $\omega_{de}$, we have
\begin{equation}
Om(z)=\Omega_{m0}+\left(1-\Omega_{m0}\right)\frac{(1+z)^{3(1+\omega_{de})}-1}{(1+z)^3-1}.
\end{equation}
The $Om(z)$ parameter is constant for $\Lambda$CDM with $\omega_{de}=-1$, increases from a constant positive value in the past to some higher positive value at $z=0$ for quintessence models with $\omega_{de}>-1$ whereas for phantom models, the  $Om(z)$ parameter decreases from a constant positive value in the past to small values close to zero at the present epoch. One may consider that the $Om(z)$ parameter may be  zero for $\Lambda$CDM universe for a vanishingly small contribution of the matter energy density \cite{Shahalam/2015,Sharma/2020}. It may be conjectured that, the deviation of $Om(z)$ from $\Omega_{m0}$ in certain dark energy models may represent a perturbation that comes from the effect rather than the matter \cite{Zhai2013}. In Fig. 9, the reconstructed $Om(z)$ parameter is shown as a function of redshift for the best fit data. The $Om(z)$ parameter is observed to decrease from positive values in the past to small positive values close to zero at the present epoch.
\begin{figure}[t!]
\centering
\includegraphics[width=7cm,height= 6cm,angle=0]{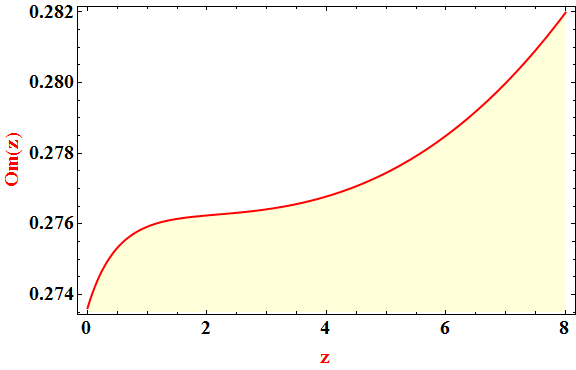} 
\caption{\bf{The plot of Om parameter $Om(z)$ over red shift $z$.}}
\end{figure}
\section{Concluding remarks}
In the present work, we have investigated a bulk viscosity dominated accelerating expansion of a spatially homogeneous but anisotropic Universe from view point of some kinematic parameters such as the Hubble parameter, deceleration parameter and jerk parameter. Basing upon a non-parametric assumptions of the kinematic parameters, we reconstructed the jerk parameter and the EoS parameter from a $\chi^2$ minimisation technique using 46 data points for observational $H(z)$ values in the redshift range $0\leq z \leq 2.36$. Also, we have carried out an analysis from the Pantheon compilation data from 1048 SNIa apparent magnitude measurements including 276 SNIa in the  low-redshift range $0.3\leq z \leq 0.65$.\\

We summarize here the main features of the present work. In our derived model, the Universe undergoes a smooth transition from a decelerated phase of expansion to an accelerated phase at a transition redshift $z_{t} = 0.73$. The extracted value of the transition redshift is compatible to some recent constraints \cite{Farooq2013, Busca2013, Lu2011, Yang2019, Capo2014}.  The transitioning Universe, we have obtained  in the present work has a deceleration parameter $q_0=-0.59$ at the present epoch. This value is quite closer to that constrained by Capozziello et al. \cite{Capo2020}. From a joint analysis of the $H(z)$ data and the Pantheon compilation data, the Hubble parameter $H_0$ is obtained to be $69.36 \pm 1.42 ~km~s^{-1}Mpc^{-1}$ which sits in a range middle to the observed Hubble tension. From the constrained value of the Hubble parameter, we obtained the age of Universe as $13.79$ Gyrs in close conformity with the Planck results \cite{Planck2020}. As a test to our derived model, we have reconstructed the jerk parameter and $Om(z)$ parameter to test whether our model deviates from the concordance $\Lambda$CDM model. Even though, the reconstructed jerk parameter at the present epoch for the best fit curve from the observational Pantheon compilation data is closer to the  concordance $\Lambda$CDM value $j_0=1$, the joint analysis predicts a departure from $\Lambda$CDM model. This deviation of the jerk parameter indicates that bulk viscous Universe is not competing with $\Lambda$CDM Universe. In the recent time, it has been widely discussed in the literature whether cosmological model other than $\Lambda$CDM may solve the $H_{0}$ tension (see Refs. \cite{Verde/2019,Poulin/2019,Nunes/2018,Yang/2018,Valentino/2019,Vagnozzi/2019,Vagnozzi/2020,Haridasu/2020,Valentino/2020a,Valentino/2020b}). It may be remarked here that, a departure from the $\Lambda$CDM model may hint for an interaction between the dark energy and dark matter components. The reconstruction of the $Om(z)$ parameter for the present bulk viscous model shows that it varies from positive values in the past to small positive values at present epoch. It may be inferred from the behaviour of the $Om(z)$ parameter that our model may favour a phantom phase at least at the present epoch. \\

As a final remark, we say that, the bulk viscous anisotropic Universe as constructed in the present work signals a possible departure from the concordance $\Lambda$CDM model while providing  reasonable estimates of the age of the Universe, the deceleration parameter, the Hubble parameter. In view of this, more involved attention should be paid to this feature through the inclusion of more recent observational data may be from the recently discovered gravitational waves.


\begin{thebibliography}{99}
\bibitem{Riess98} A. G. Riess et al., \textit{Astron. J.}, \textbf{116}, 1009 (1998).

\bibitem{Perlmutter99} S. Perlmutter et al., \textit{Astrophys. J.}, \textbf{517}, 565 (1999).

\bibitem{Ade16} P. A. R. Ade et al., \textit{Astron. \& Astrophys.}, \textbf{594} A14 (2016).

\bibitem{Ma99} C. P. Ma, R. R. Caldwell, P. Bode, L. Wang, \textit{Astrophys. J.}, \textbf{521}, L1 (1999).

\bibitem{Sahni03} V. Sahni, Y. Shtanov, \textit{J. Cosmol. Astropart. Phys.}, \textbf{11}, 014 (2019).

\bibitem{Wei08} H. Wei, R. G. Cai, \textit{Phys. Lett. B}, \textbf{660}, 113 (2008)

\bibitem{Caplar13} N. Caplar, H. Stefancic, \textit{Phys. Rev. D}, \textbf{87}, 023510 (2013).

\bibitem{Mishra15} B.Mishra, S.K. Tripathy, \textit{Mod. Phys. Lett. A}, \textbf{30}, 1550175 (2015).

\bibitem{Mishra17} B. Mishra, P.P. Ray, S.K.J. Pacif, \textit{Eur. Phys. J. Plus}, \textbf{132}, 429 (2017).

\bibitem{Mishra18a} B. Mishra, P.P. Ray, S.K.J. Pacif, \textit{Adv. High Energy Phys.}, \textbf{2018}, 6306848 (2018).

\bibitem{Farnes18} J.S. Farnes, \textit{Astron. Astrophys}, \textbf{620}, A92 (2018).

\bibitem{Mishra18b} B. Mishra, S.K. Tripathy, P.P. Ray, \textit{Astrophys. Space Sci.}, \textbf{363} 86 (2018)

\bibitem{Ray19} P.P. Ray, B.Mishra, S.K. Tripathy, \textit{Int. J. Mod. Phys. D}, \textbf{28}, 1950093 (2019).

\bibitem{Mishra19a} B. Mishra, P.P. Ray, S.K. Tripathy, K. Bamba, \textit{Mod. Phys. Lett. A}, \textbf{34}, 1950217 (2019).

\bibitem{Yadav/2011} A. K. Yadav, \textit{Astrophys. Space Sci.}, \textbf{335}, 565 (2011).

\bibitem{Kumar/2011mpla} S. Kumar, A. K. Yadav, \textit{Mod. Phys. Lett. A} \textbf{26}, 647 (2011).

\bibitem{Amirhashchi/2011plb} H. Amirhashchi, \textit{Phys. Lett. B} \textbf{697}, 429 (2011).

\bibitem{Yadav/2012} A. K. Yadav et al., \textit{Eur. Phys. J. Plus} \textbf{127}, 127 (2012).

\bibitem{Goswami/2015} G. K. Goswami, M. Mishra, A. K. Yadav, \textit{Int. J. Theor. Phys.} \textbf{54}, 315 (2015).

\bibitem{Goswami/2016} G. K. Goswami, A. K. Yadav, R. N. Dewangan, A. Pradhan, \textit{Astrophys. Space Sci.} \textbf{361}, 47 (2016).

\bibitem{Goswami/2016a} G. K. Goswami, R. N. Dewangan, A. K. Yadav, \textit{Astrophys. Space Sci.} \textbf{361}, 119
(2016).

\bibitem{Goswami/2016b} G. K. Goswami, A. K. Yadav, R. N. Dewangan, \textit{Int. J. Theor. Phys.} \textbf{55}, 4651
(2016).

\bibitem{Singla/2020} N. Singla, A. K. Yadav, M. K. Gupta, G. K. Goswami R. Prasad, \textit{Mod. Phys. Lett. A}, 2050174 (2020).

\bibitem{SKT2015} S. K. Tripathy, D. Behera and B. Mishra, \textit{Eur. Phys. J. C}, \textbf{75}, 149 (2015).

\bibitem{SKT2020} S. K. Tripathy, S. K. Pradhan, Z. Naik, D. Behera and B. Mishra, \textit{Phys. Dark Univ.}, \textbf{30}, 100722 (2020).

\bibitem{Cicoli20} M. Cicoli, G. Dibitetto, F. G. Pedro, \textit{Phys. Rev. D}, \textbf{101},103524 (2020).

\bibitem{Chakraborty20} S. Chakraborty, S. Mishra, S. Chakraborty, \textit{Eur. Phys. J. C}\textbf{80}, 852 (2020).

\bibitem{Alestas20} G. Alestas, L. Kazantzidis, L. Perivolaropoulos, \textit{Phys. Rev.D.}, \textbf{101}, 123516 (2020).

\bibitem{Dantas20} J. D. Dantas,  J.J. Rodrigues, \textit{ Eur. Phys. J. C}, \textbf{80}, 666 (2020).

\bibitem{Cheng20} G. Cheng, Yin-Zhe Ma, F. Wu, J. Zhang, X. Chen, \textit{Phys. Rev. D}, \textbf{102}, 043517 (2020).

\bibitem{Paliathanasis20} A. Paliathanasis, G. Leon, \textit{ Eur. Phys. J. C}, \textbf{80} , 589 (2020).

\bibitem{Lin20} K. Lin, W. Qian, \textit{Eur. Phys. J. C}, \textbf{80}, 561 (2020).

\bibitem{Padmanabhan87} T. Padmanabhan, S. M. Chitre, \textit{Phys. Lett. A}, \textbf{120}, 433 (1987).

\bibitem{Zimdahl01} W. Zimdahl, D.J. Schwarz, A.B. Balakin, D. Pavon, \textit{Phys. Rev. D}, \textbf{64}, 063501 (2001).

\bibitem{Cataldo05} M. Cataldo, N. Cruz, S. Lepe, \textit{Phys. Lett. B}, \textbf{619}, 2 (2005).

\bibitem{Brevik05} I. Brevik, O. Gorbunova, \textit{Gen. Rel. Grav}, \textbf{37}, 2039 (2005).

\bibitem{Brevik11} I. Brevik, E. Elizalde, S. Nojiri, S. D. Odintsov, \textit{Phys. Rev. D}, \textbf{84}, 103508 (2011).

\bibitem{Brevik12} I. Brevik, R. Myrzakulov, S. Nojiri, S. D. Odintsov, \textit{Phys. Rev. D}, \textbf{86}, 063007 (2012).

\bibitem{Velten13} H. Velten, J. Wang, X. Meng, \textit{Phys. Rev. D}, \textbf{88}, 123504 (2013).

\bibitem{Kumar13} S. Kumar, \textit{Grav. Cosmo.}, \textbf{19}, 284 (2013).

\bibitem{Amirhashchi14} H. Amirhashchi, \textit{Astrophys. Space Sci.}, \textbf{351} 641 (2014).

\bibitem{Rezaei17} M. Rezaei, M. Malekjani, \textit{Phys. Rev. D}, \textbf{96}, 063519 (2017).

\bibitem{Wang17} D. Wang, Y.J. Yan, X.H. Meng, \textit{Eur. Phys. J. C}, \textbf{77}, 660 (2017).

\bibitem{Amirhashchi17} H. Amirhashchi, \textit{Phys. Rev. D}, \textbf{96}, 123507 (2017).

\bibitem{Atreya19} A. Atreya, J. R. Bhatt, A. K. Mishra, \textit{J. Cosmol. Astropart. Phys.}, \textbf{02}, 045 (2019).

\bibitem{Mishra19b} B. Mishra, P. P. Ray, R. Myrzakulov, \textit{Eur. Phys. J. C}, \textbf{79}, 34 (2019).

\bibitem{Silva19} W.J.C. da Silva, R. Silva,  \textit{J. Cosmol. Astropart. Phys.} \textbf{05}, 036 (2019).

\bibitem{Odintsov20} S. D. Odintsov, D.S. Gómez, G. S. Sharov, \textit{Phys. Rev. D}, \textbf{101}, 044010 (2020).

\bibitem{Hinshaw03} G. Hinshaw et al, \textit{Astrophys. J. Suppl.}, \textbf{148}, 135 (2003).

\bibitem{Hinshaw07} G. Hinshaw et al, \textit{Astrophys. J. Suppl.}, \textbf{170}, 288 (2007).

\bibitem{Biswas/2019} P. Biswas, P. Roy, R. Biswas, arXiv: 1908.00408 [gr-qc] (2019).

\bibitem{Scolnic/2018} D. M. Scolnic et al, \textit{Astrophys. J}, \textbf{859}, 101 (2018).

\bibitem{Riess2016} A.G. Riess, et al., \textit{Astrophys. J.}, \textbf{826 (1)}, 56 (2016). arXiv:1604.01424

\bibitem{Riess2018} A.G. Riess, et al., \textit{Astrophys. J.}, \textbf{861 (2)}, 126 (2018). arXiv:1804.10655

\bibitem{Riess2019} A. G. Riess, S. Casertano, W. Yuan, L. M. Macri and D. Scolnic, \textit{Astrophys. J.}, \textbf{876}, 85 (2019).

\bibitem{Planck2020} N. Aghanim et al., \textit{Planck Collaboration}, \textit{Astron. Astrophys}, \textbf{641}, A6 (2020). arXiv:1807.06209

\bibitem{Reid2019} M.J. Reid, D.W. Pesce, A.G. Riess, \textit{Astrophys. J.}, \textbf{886 (2)}, L27 (2019), arXiv:
1908.05625.

\bibitem{Wong2019} K.C. Wong, et al., \textit{H0LiCOW Collaboration}, arXiv:1907.04869.

\bibitem{LIGO2017} LIGO Scientific, VIRGO, 1M2H, Dark Energy Camera GW-E, DES, DLT40, Las Cumbres Observatory, \textit{Nature}, \textbf{551}, 85 (2017).arXiv;1710.05835

\bibitem{Freedman2019} W. L. Freedman, B. F. Madore, D. Hatt, T. J. Hyot et al., arXiv:1907.05922.

\bibitem{Yadav/2021} A. K. Yadav, A. M. Alshehri, N. Ahmad,G. K. Goswami, M. Kumar , \textit{Phys. Dark Univ.}, \textbf{31}, 100738 (2021)

\bibitem{Prasad/2020Pramana} R. Prasad, L. K. Gupta, G. K. Goswami, A. K. Yadav, \textit{Pramana J. Phys.} \textbf{94}, 135 (2020).

\bibitem{Prasad/2020} R, Prasad, A. K Yadav \& A. K. Yadav, \textit{Euro. Phys. J. Plus}, \textbf{135}, 297 (2020)

\bibitem{Lambiase2019} G. Lambiase, S. Mohanty, A. Narang, P. Parashari, \textit{Eur. Phys. J. C}, \textbf{79}, 141 (2019).

\bibitem{Sola2020} J. S. Peracaula, A.G. Valent, J. De C. Perez, C.M. Pulido, arXiv:2006.04273.

\bibitem{Adhikari2020} S. Adhikari and D. Huterer, \textit{Phys. Dark. Univ.}, \textbf{28}, 100539 (2020).

\bibitem{Bernal2016} J. L. Bernal,  L. Verde,  and A. G. Riess, \textit{J. Cosmol. Astropart. Phys.}, \textbf{1610}, 019 (2016).

\bibitem{Mortsell2018} E. Mortsell, S. Dhawan, \textit{J. Cosmol. Astropart. Phys.}, \textbf{1809}, 025 (2018).

\bibitem{Farooq2013} O. Farooq and B. Ratra, \textit{Astrophys. J. Lett.}, \textbf{766}, L7 (2013).

\bibitem{Busca2013} N. Busca, \textit{Astron. Astrophys.}, \textbf{552}, A96 (2013).

\bibitem{Lu2011} J.Lu, L. Xu and M. Liu, \textit{Phys. Lett. B}, \textbf{699}, 246 (2011).

\bibitem{Yang2019} Y. Yang and Y. Gong, arXiv:1912.07375.

\bibitem{Capo2014} S. Capozziello, O. Farooq, O. Luongo and B. Ratra, \textit{Phys. Rev. D}, \textbf{90}, 044016 (2014).

\bibitem{Camarena2020} D. Camarena and V. Marra, \textit{Phys. Rev. Res.}, \textbf{2}, 013028 (2020).

\bibitem{Capo2020}S. Capozziello, R. D'Agostino and O. Luongo, \textit{Mon. Not. Roy. Astron. Soc.}, \textbf{494}, 2576 (2018).

\bibitem{Bentabol/2013} B. M. Bentabol, J. M. Bentabol, J. Cepa, \textit{J. Cosmol. Astropart. Phys.} \textbf{02}, 015 (2013).

\bibitem{Goswami/2020} G. K. Goswami, M. Mishra, A. K. Yadav \& A. Pradhan, \textit{Mod. Phys. Lett. A}, \textbf{35}, 2050086 (2020).

\bibitem{Sahni2003} V. Sahni, T. D. Saini, A. A. Starobinsky and U. Alam, \textit{JETP Lett.}, \textbf{77}, 201 (2003).

\bibitem{Alam2003} U. Alam, V. Sahni, T. D. Saini,  A. A. Starobinsky, \textit{Mon. Not. Roy. Astron. Soc.}, \textbf{344}, 1057 (2003).

\bibitem{Zhai2013} Z. X. Zhai, M. J. Zhang, Z. S> Zhang, X. M. Liu, T. J. Zhang, \textit{Phys. Lett. B}, \textbf{727}, 8 (2013).

\bibitem{Mamon2018} A. A. Mamon and K. Bamba, \textit{Eur. Phys. J. C}, \textbf{78}, 862 (2018).

 
\bibitem{Mukherjee2017} A. Mukherjee and N. Banerjee, \textit{Class. Quant. Gravit.}, \textbf{ 4}, 035016 (2017).

\bibitem{Sahni2008} V. Sahni, A. Shafieloo, A. A. Starobinsky, \textit{Phys. Rev. D}, \textbf{78}, 103502 (2008).

\bibitem{Sahni2014} V. Sahni, A. Shafieloo, A. A. Starobinsky, \textit{Astrophys. J.}, \textbf{793}, L40 (2014).

\bibitem{Shahalam/2015} M. Shahalam, S. Sami and A. Agarwal, \textit{Mon. Not. R Astron. Soc.}, \textbf{448}, 2948 (2015).

\bibitem{Sharma/2020} L. K. Sharma, B. K. Singh, A. K. Yadav, \textit{Int. J. Geom. Meth. Mod. Phys.},\textbf{17}, 2050111 (2020).

\bibitem{Verde/2019} L. Verde, T. Treu, A.G. Riess, {\it Nature Astronomy} {\bf 3}, 891 (2019).

\bibitem{Poulin/2019} V. Poulin, T. L. Smith, T. Karwal, M. Kamionkowski, {\it Phys. Rev. Lett.} {\bf 122}, 221301 (2019). 

\bibitem{Nunes/2018} R. C. Nunes, {\it J. Cosmol. Astrop. Phys.} {\bf 05}, 052 (2018).

\bibitem{Yang/2018} W. Yang et al., {\it J. Cosmol. Astrop. Phys.} {\bf 09}, 019 (2018).

\bibitem{Valentino/2019} E. Di Valentino, A. Melchiorri, O. Mena, S. Vagnozzi, arXiv: 1908.04281

\bibitem{Vagnozzi/2019} S. Vagnozzi, arXiv: 1907.07569

\bibitem{Vagnozzi/2020} S. Vagnozzi, E. Di Valentino, S. Gariazzo, A. Melchiorri, O. Mena, J. Silk, arXiv: 2010.02230

\bibitem{Haridasu/2020} B. S. Haridasu, M. Viel, {\it Mon. Not. Roy. Astron. Soc.} {\bf 497}, 2 (2020). 

\bibitem{Valentino/2020a} E. Di Valentino, arXiv: 2011.00246

\bibitem{Valentino/2020b} E. Di Valentino, A. Melchiorri, O. Mena, S. Pan, W. Yang, arXiv: 2011.00283



\end{thebibliography}
\end{document}